\documentclass[useAMS,usenatbib]{mnras}

\usepackage{graphicx}
\usepackage{microtype}
\bibpunct{(}{)}{;}{a}{}{,} 
\usepackage{amssymb}
\usepackage{hyperref}  
\def\link_col{blue}
\hypersetup{colorlinks,citecolor=\link_col,filecolor=\link_col,linkcolor=\link_col,urlcolor=\link_col}
\usepackage{xspace}
\usepackage{url}
\usepackage{lscape}   
\usepackage{multicol} 
\usepackage{verbatim} 
\usepackage{multirow} 
\usepackage{rotating} 
\usepackage{wasysym}
\usepackage{journal_names}

\def\pasa{\ref@jnl{PASA}}

\newcommand\etal{\emph{et al.}\xspace}
\newcommand\halpha{\mbox{H$\alpha$\xspace}~}
\newcommand\emunits{\mbox{pc~cm$^{-6}$}~}
\newcommand\cmcube{\mbox{cm$^{-3}$}\xspace}
\newcommand\alphaem{\mbox{$\alpha_{\rm EM}$}}

\title[Absorption of point sources as a probe of the clumpy-WIM]{A new method to probe the thermal electron content of the Galaxy through spectral analysis of background sources}

\author[Jones \etal]{Jones, D. I.$^1$, Igoshev, A.~P.$^1$ \& Haverkorn, M.$^{1,2}$\thanks{E-mail corresponding author: m.haverkorn@astro.ru.nl}\\
$^{1}$Department of Astrophysics/IMAPP, Radboud University, PO Box 9010, 6500 GL Nijmegen, The Netherlands. \\
$^{2}$Leiden Observatory, Leiden University, PO Box 9513, 2300 RA Leiden, The Netherlands. \\
}
\begin{document}

\date{Accepted \today. Received \today; in original form \today}

\pagerange{\pageref{firstpage}--\pageref{lastpage}} \pubyear{2014}

\maketitle

\label{firstpage}

\begin{abstract}
We present a new method for probing the thermal electron content of the Galaxy by spectral analysis of background point sources in the absorption-only limit to the radiative transfer equation.
In this limit, calculating the spectral index, $\alpha$, of these sources using a natural logarithm results in an additive factor, which we denote $\alphaem$, resulting from the absorption of radiation due to the Galactic thermal electron population.
We find that this effect is  important at very low frequencies ($\nu\lesssim200$~MHz), and that the frequency spacing is critical.

We model this effect by calculating the emission measure across the sky. A (smooth) thermal electron model for the Galaxy does not fit the observed emission measure distribution, but a simple, cloud-based model to represent the clumpy nature of the warm interstellar medium does. This model statistically reproduces the Galactic emission measure distribution as obtained independently from \halpha data well.

We find that at the lowest frequencies ($\sim10-50$~MHz), the observed spectral index for a large segment of the Galaxy below Galactic latitudes of $\lesssim15^\circ$ could be changed significantly (i.e., $\alphaem\gtrsim0.1$).

This method therefore provides a correction to low-frequency spectral index measurements of extragalactic sources, and provides a sensitive probe of the thermal electron distribution of the Galaxy using current and next-generation low-frequency radio telescopes. 

We show that this effect should be robustly detectable individually in the strongest sources, and statistically in source samples at a level of $\alphaem\gtrsim0.18,0.06$, and 0.02 for source densities of 10, 100 and 1,000 sources per square degree.

\end{abstract}


\section{Introduction}
A knowledge of the interactions between the phases that make up the interstellar medium (ISM) is fundamental in our attempt to understand the structure and evolution of our Galaxy.
%
%
The importance of thermal electrons, denoted $n_e$, in the determination of the Galactic magnetic field is twofold.
Firstly, it is crucial for the determination of distances to pulsars through the dispersion measure, DM~$\equiv\int n_eds$ in units of pc~\cmcube \citep{LorimerKramer04}.
Secondly, it is required for an accurate determination of the strength and structure of the Galactic magnetic field, as $n_e$ enters into the determination of rotation measure (which is defined as RM~$\equiv0.81\int n_e\mathbf{B}\cdot d\mathbf{s}$~rad~m$^{-2}$, with $n_e$ in cm$^{-3}$, $\mathbf{B}$ in micro-Gauss, and $d\mathbf{s}$ in parsecs).
Thus the RM-grid method of exploring the magnetic field of our Galaxy  (e.g., \citealt{Farnes2014}), as well as the determination of small-scale structure through the RM synthesis technique (\citealt{Brentjens2005}) depend on a good knowledge of $n_e$. Although the sign of RM exclusively depends on the direction of the line-of-sight magnetic field, the electron density is needed for estimates of the magnitude of these magnetic fields.

In this paper, we explore the relationship between the spectral index of background point sources and the thermal electron content of the Galaxy.
Specifically, we show that using a natural logarithm in the calculation of the spectral index, assuming absorption-only, results in a additive factor to the spectral index that is proportional to the emission measure (EM). This is equivalent to a multiplicative factor to flux.
We show that this parameter becomes significant at the lowest radio frequencies, such that so-called ``in-band'' spectral analysis at the lowest frequencies for the next-generation low-frequency radio telescopes such as the Low Frequency ARray (LOFAR) and the Murchison Widefield Array (MWA), might be severely affected by this phenomenon.
This method can thus also be used to obtain an estimate of the thermal electron content of the Galaxy.
Furthermore, we show that a simple smooth model for $n_e$, as obtained from pulsar measurements (e.g, the modified Taylor-Cordes model; \citet{Schnitzeler2012} after \citealt{Taylor1993}), is inadequate to reproduce the distribution in EM  as obtained from \halpha observations \citep{Hill2008}.
We show that a simple ``clumpy'' model, based on that for pulsar DMs \citep{Nelemans1997} can re-produce the observed EM distribution, as well as render the effect derived here an even more important factor in low-frequency radio observations.
Finally, we show that typical source densities as obtained with current and future low-frequency radio telescopes such as the LOFAR, MWA and eventually the Square Kilometre Array (SKA) imply that this effect should be statistically detected, even at high Galactic latitudes. In addition, detection in individual sources should be possible for the strongest sources with currently planned or ongoing low-frequency surveys.

\section{Spectral Curvature at Low Frequencies}
The solution to the equation of radiative transfer assuming absorption-only relates the intrinsic flux density, $S^{\rm int}_\nu$, at frequency $\nu$ to the observed flux density, $S^{\rm obs}_\nu$, modified by an exponential term representing the optical depth, $\tau_\nu$, of the photon (e.g., \citealt{Rybicki}):
\begin{equation}\label{eq:PL}
	S^{\rm obs}_\nu = S^{\rm int}_\nu e^{-\tau_\nu}.
\end{equation}
A value of $\tau_\nu\gtrsim1$ will result in the absorption of the intrinsic flux density, which may alter the observed flux density considerably.
Defining the spectral index using natural logarithms (instead of the canonical $\log_{10}$ definition) leverages the information added by the optical depth term appearing in Equation~\ref{eq:PL} in as simple a manner as possible:
\begin{eqnarray}\label{eq:alpha}
	\alpha_{\rm tot} & \equiv & d\ln(S^{\rm obs}_{\nu})/d\ln(\nu) \\
	& \approx & \frac{\ln([S^{\rm int}_{\nu_1}e^{-\tau_{\nu_1}}]/[S^{\rm int}_{\nu_2}e^{-\tau_{\nu_2}}])}{\ln(\nu_1/\nu_2)} \nonumber \\
	& = & \frac{\ln(S^{\rm int}_{\nu_1}/S^{\rm int}_{\nu_2})}{\ln(\nu_1/\nu_2)}+\frac{(\tau_{\nu_2}-\tau_{\nu_1})}{\ln(\nu_1/\nu_2)} \label{eq:alpha2} \\
	& \equiv &\alpha_{\rm int} + \alphaem \nonumber,
\end{eqnarray}
where $\nu_{1,2}$ refers to the frequencies at which the spectral index is calculated and $\alpha_{\rm int}$ is the spectral index intrinsic to the source. We denote $\alphaem$, which we defined above, as the ``spectral thermal absorption parameter'', or STAP.
The optical depth is the integral over the line-of-sight $ds$ of the opacity $\kappa_\nu$, which is a function of the square of the electron density (e.g. \citealt{Peterson2002}):
\begin{equation}
	\frac{\kappa_\nu}{\textrm{cm}^{-1}}=1.78\times 10^{-20} g^\nu_{\rm ff}\left(\frac{n_e}{\textrm{cm}^{-3}}\right)^2 \left(\frac{\nu}{\textrm{GHz}}\right)^{-2}\left(\frac{T_e}{\textrm{K}}\right)^{-3/2},
\end{equation}
where $g_{\rm ff}^\nu$ is the Gaunt factor and is of order unity:
\begin{equation}
g^\nu_{\rm ff} = 10.6 + 1.9\log\left(\frac{T_e}{\textrm{K}}\right) - 11.34\log\left(\frac{\nu}{\textrm{GHz}}\right).
\end{equation}

Hence, one can obtain a relation between the optical depth and emission measure EM~$\equiv\int n_e^2ds$~\emunits):
\begin{equation}
\tau_\nu = 0.055 \overline{g_{\rm ff}^\nu} \left( \frac{\nu}{\textrm{GHz}}\right)^{-2}\left(\frac{T_e}{\textrm{K}}\right)^{-3/2} \left( \frac{\textrm{EM}}{\textrm{pc cm}^{-6}} \right),
\end{equation}
where $\overline{g_{\rm ff}^\nu}$ is the average Gaunt factor. Thus STAP is a function of EM.
Putting this all together, one obtains a simple relation for how the spectral index changes, $\alphaem$, as a function of observing frequencies and thermal electron content along a line-of-sight:
$$
\alphaem=5.5\times10^{-2} \left(\frac{\textrm{EM}}{\textrm{pc cm}^{-6}}\right)\left(\frac{\textrm{T}_e}{\textrm{K}} \right)^{-3/2} \frac{1}{\ln(\nu_1 / \nu_2)}
$$
\begin{equation} \label{eq:STAP}
\times \left[g_{\rm ff}^{\nu_2}\left(\frac{\nu_2}{\textrm{GHz}}\right)^{-2} -  g_{\rm ff}^{\nu_1} \left(\frac{\nu_1}{\textrm{GHz}}\right)^{-2} \right]
\end{equation}

where the factor of $5.5\times10^{-2}$ accounts for the unit conversion of EM to opacity $\kappa_\nu$, and $\nu_{1,2}$ are the frequencies in Hz at which the spectral index is calculated. 
$T_e$ is the electron temperature and we assume a value of $T_e=10^4$~K throughout this paper.
The mean thermal electron density of the {\it intergalactic} medium is $\sim2.1\times 10^{-7}$~cm$^{-3}$ \citep{Inoue2014}, which makes its contribution negligible compared to that of the Galaxy, so that the absorption of background sources at low frequencies is due solely to the Galactic population of thermal electrons.

\begin{figure*}
\centering
\includegraphics[width=0.5\textwidth]{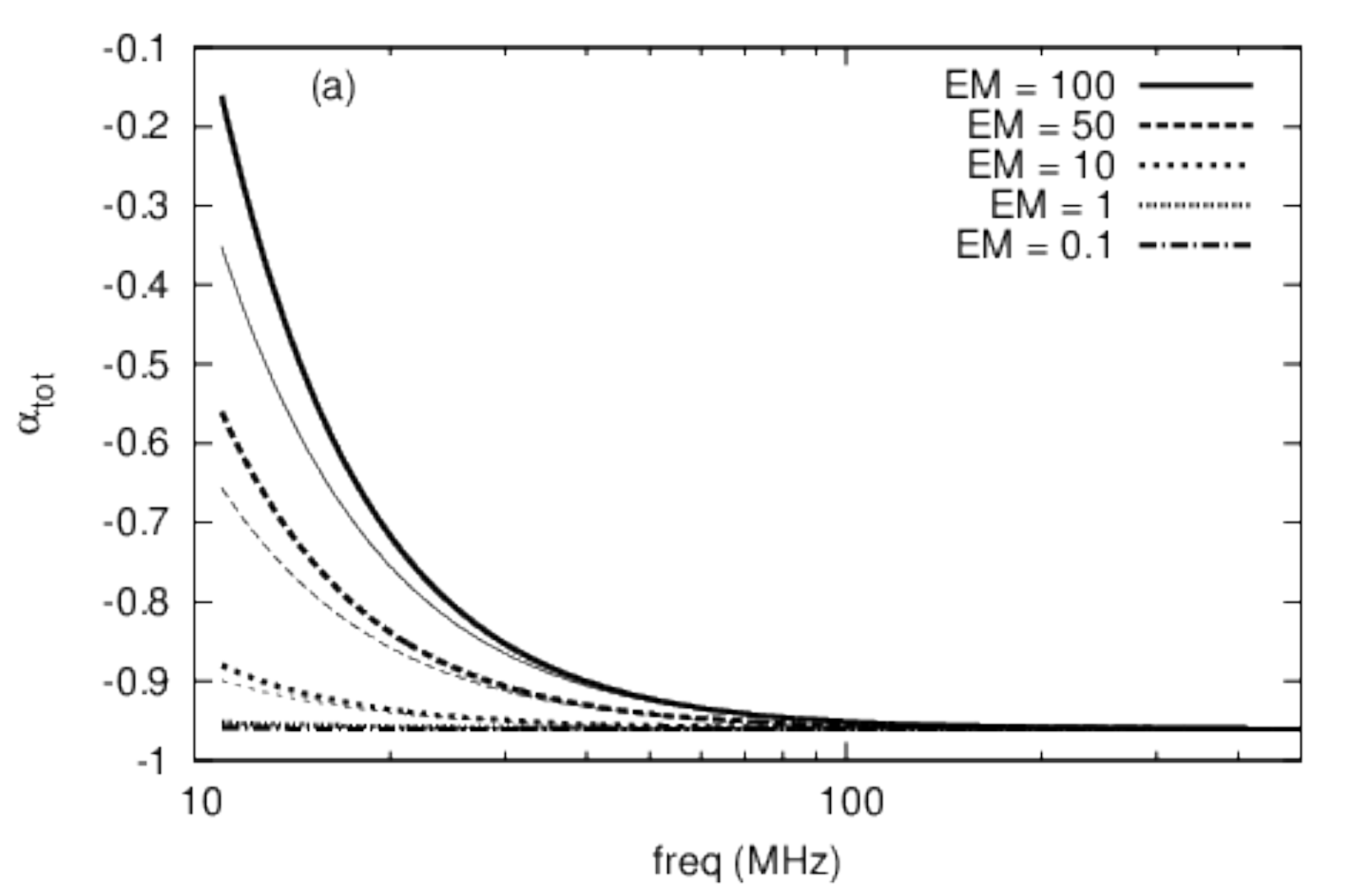}\includegraphics[width=0.5\textwidth]{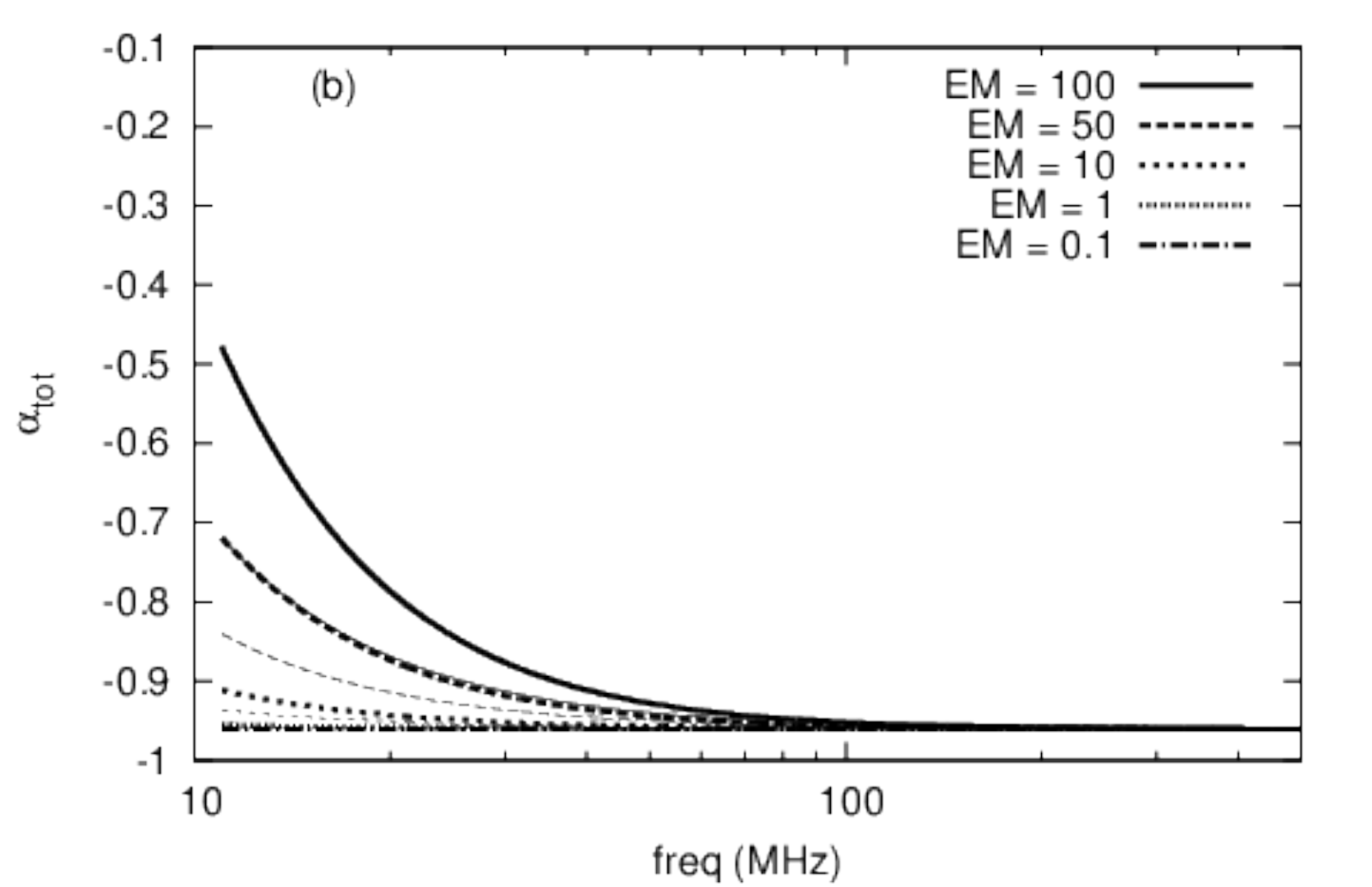}
\caption{Plots of the {\it total} spectral index, $\alpha_{\rm tot}$, illustrating the change from the intrinsic spectral index ($\alpha_{\rm int}$) that the STAP (i.e., $\alphaem$) has as a function of frequency from 10 to 500~MHz for a number of different frequency widths.
A canonical intrinsic spectral index of $\alpha_{\rm int}=-0.96$ was used, and hence the STAP is any deviation away from a horizontal line.
Plot (a) shows the STAP for separations of both 1 (thick) and 5~MHz (thin) lines as labelled, whilst plot (b) shows the same for separations in frequency of 10 and 50~MHz (thick and thin lines, respectively).
The plots both show the same frequency and spectral index range, so as to illustrate the difference in the change between 1 and 5~MHz and 10 and 50~MHz steps.
}
\label{fig:STAP}
\end{figure*}

To study the effect of absorption, we first took a nominal source taken to have an (arbitrary) intrinsic spectral index of $\alpha_{\rm int}=-0.96$. 
We then calculated the spectral index for this source using Equation~\ref{eq:STAP} for a range of plausible EMs for the warm ionised medium (WIM); namely 0.1, 1.0, 10, 50 and 100~\emunits.
Figures~\ref{fig:STAP}(a) and (b) show the results for steps in frequency of 1 and 5~MHz (thick and thin lines of Figure~\ref{fig:STAP}(a), respectively), and 10 and 50~MHz (thick and thin lines of Figure~\ref{fig:STAP}(b), respectively).
These figures show a number of things.
Firstly, for reasonable (see below) EMs, such a change in spectral index may be detectable, and secondly, the spacing between spectral index calculations at low frequencies matters.
%
%
If a source is strong enough such that high signal-to-noise can be obtained in small (e.g., $\sim1$~MHz) channels, then this effect can be exploited to obtain information about the thermal electron content of the Galaxy even at relatively high latitudes (c.f. Figure~\ref{fig:EM}).

We must note at this juncture that such effects are well known: at very-low-frequencies, the turnover of the Galactic synchrotron spectrum has been used to infer properties about the WIM as well \citep{Peterson2002}.
Additionally, \citet{Kassim1989} explored absorption towards the envelopes of H{\sc ii} regions because on large scales in our Galaxy, it was thought that the average density of thermal electrons along a particular line of sight is inadequate to produce significant amounts of absorption.
Using an average thermal electron density of $n_e\approx0.03$~\cmcube at a frequencies of 150 and 330~MHz, this equates to optical depths of $\tau_{150}\sim10^{-4}$ and $\tau_{330}\sim10^{-3}$ (i.e., $\tau_\nu\ll1$), this would indeed seem to be the case.
However, given Figures~\ref{fig:STAP}(a) and (b), as well as the fact that no one has -- to the best of our knowledge -- investigated such effects in a systematic way, nor obtained the above simple-relation between the EM and spectral index, we tested a model for the distribution of thermal electrons in the Galaxy to test on a real catalogue in an attempt to explore this effect.

\subsection{The STAP, the modified-TC93 $n_e$ model and the Farnes, et~al. ``meta-catalogue''}
We have obtained the modified Taylor-Cordes (modified-TC93; \citealt{Schnitzeler2012}) model for the thermal electron distribution of the Galaxy, and the ``meta-catalogue'' of \citet{Farnes2014}.
This meta-catalogue is compiled of data from a number of radio catalogues, such as NRAO VLA Sky Survey (VLSS), AT20G, B3-VLA, GB6, NORTH6CM, Texas, and WENSS, in order to characterise the nature of polarised background galaxies for future experiments in the rotation measure grid (RM-grid; e.g., \citealt{Gaensler2009}).
%
We use the source coordinates in this catalogue to calculate the emission measure towards that particular location using the modified-TC93 model.
We integrate for a distance of 30~kpc in all directions for simplicity; doing this does not lead to significantly different results from using a more realistic integration length, such as stopping when two consecutive steps produce no change in the EM.

\begin{figure*}
\centering
\includegraphics[width=0.48\textwidth]{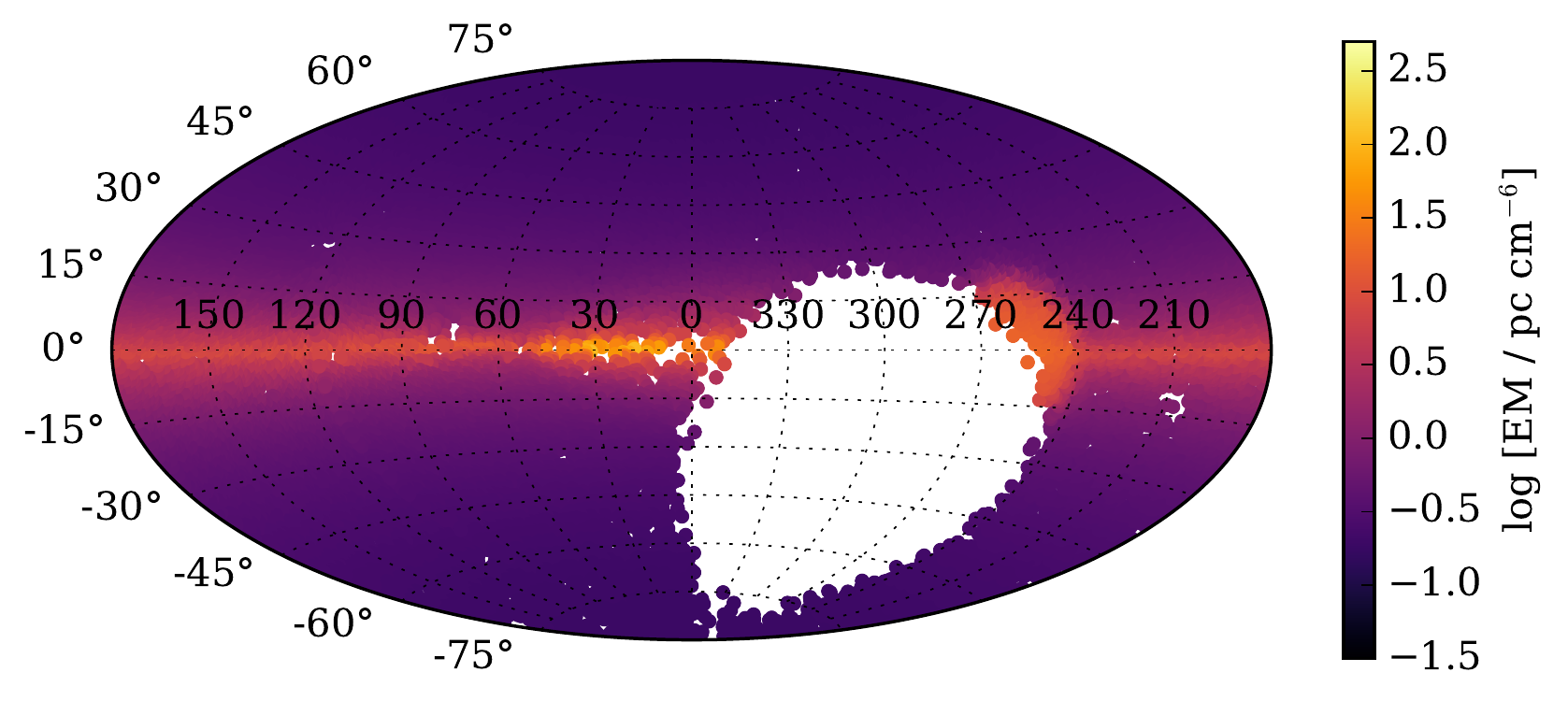}\includegraphics[width=0.48\textwidth]{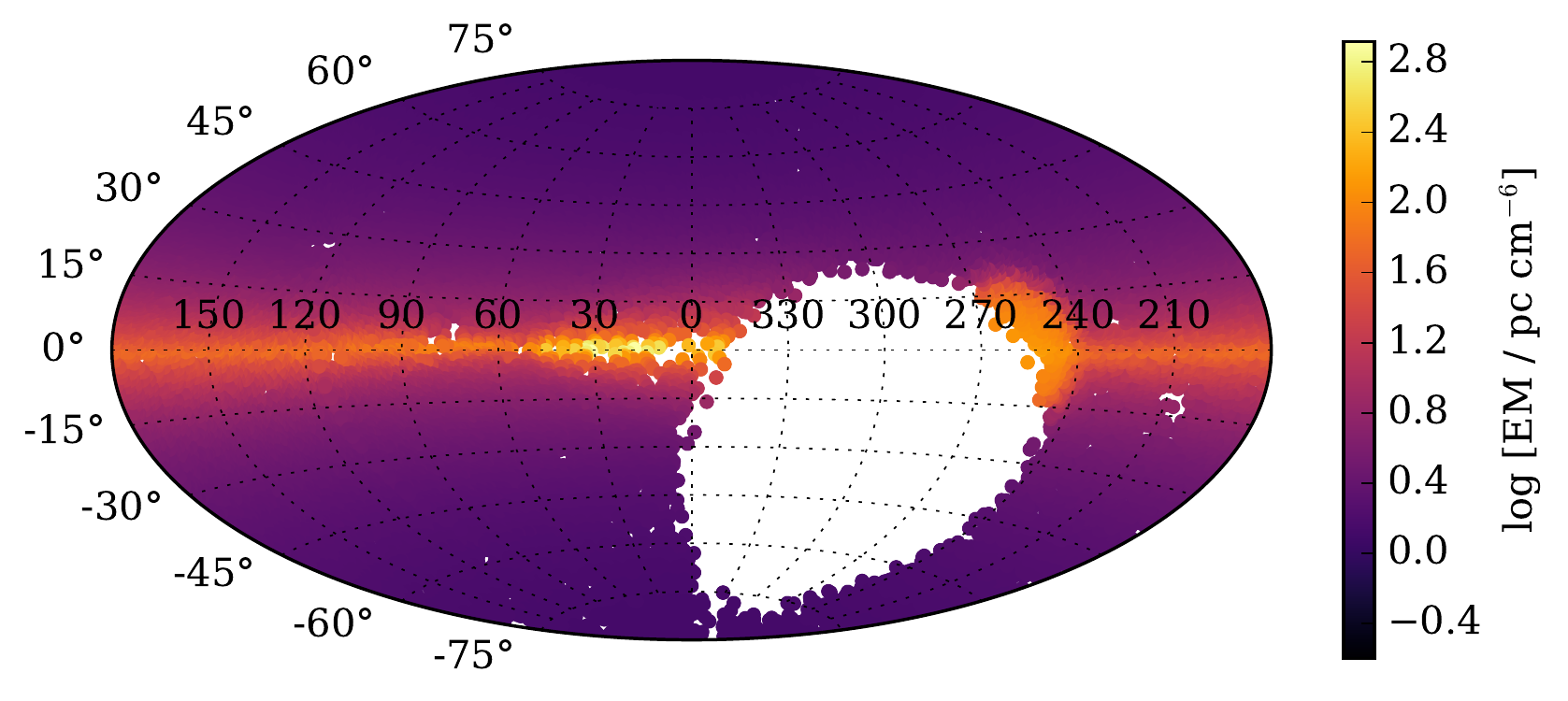}
\caption{Aitoff projection of EM for sources in the \citet{Farnes2014} meta-catalogue obtained by (a) the smooth thermal electron distribution in the Galaxy as according to the modified-TC93 model, and (b) the smooth modified TC93 model in combination with our clumpy model using a filling factor of $f=0.1$ and an (average) cloud--cloud length of $L=0.43$~kpc according to Equation~\ref{eq:EMtot}. The colour-scale is in units of pc~cm$^{-6}$ and has had a (base--10) logarithm transfer function applied to it, so as to accentuate low-EM regions.}
\label{fig:EM}
\end{figure*}

In Figure~\ref{fig:EM}(a) we show the results of this calculation.
We chose to use the EM in these plots because it is frequency independent and shows that the EM range shown in Figures~\ref{fig:STAP}(a) and (b) are reasonable.
Calculating the STAP for the data in this catalogue would not be useful, since the majority of sources are measured at relatively high frequencies (between 1 and 2~GHz). However, it is worth using such a catalogue so as to illustrate the coverage of a future catalogue from LOFAR, the MWA or SKA, which were not available at the time of writing.
Figure~\ref{fig:EM}(a) shows, using the modified-TC93 model, that below latitudes of $10-15^{\circ}$, the average EM is somewhere between 1 and 100~\emunits.
%
Figure~\ref{fig:STAP}(a) and (b) implies that a change from the intrinsic spectral index of a source measured between 10 and 20~MHz can be up to 0.5 for a EM of 100~\emunits, but ostensibly undetectable for an EM of 1~\emunits (these are represented by the thick lines in Figure~\ref{fig:STAP}(b) for a frequency difference of 10~MHz).
At a more representative LOFAR low-band frequency of 50~MHz, but still using 10~MHz wide images for spectral index calculations, the spectral index change would be 0.005 and 0.05 for EMs of 1 and 100~\emunits, respectively.
However, as we show below (c.f. Section~\ref{sec:halpha}), a smooth model, such as that used here does not represent the known distribution of EMs across the sky; the WIM is known to be clumpy \citep{Ferriere2001, Haffner2009}.

\section{The Clumpy WIM and STAP}
In this section, we model the ``clumpiness'' of the WIM by extending the model of \citet{Nelemans1997} developed to explain pulsar DM anomalies to EM.

\subsection{A simple model for a clumpy ISM}\label{sec:clumpyWIM}
As in \citet{Nelemans1997}, we define a clump as a region where the number density of electrons is enhanced by a factor, $f$, which is related to the number density, $N_c$, and radius, $R_c$, of the clump by:
\begin{equation}
f=N_c\frac{4\pi}{3}R_c^3.
\end{equation}
A source that lies behind a clump passes through it at distance $r_c\leq R_c$ from the centre of the clump will possess an emission measure of:
\begin{equation}
\textrm{EM}(r_c) = \frac{2n_e^2}{f^2}\sqrt{R_c^2-r_c^2}.
\end{equation}
Thus, the average emission measure for a collection of lines-of-sight passing through an over-dense region is given by:
\begin{equation}\label{eq:EMcl}
\langle \textrm{EM} \rangle = \int^{R_c}_{0}\frac{2n_e^2}{f^2}\sqrt{R_c^2-r^2_c}\frac{2\pi r_cdr_c}{\pi R_c^2}=\frac{4n_e^2R_c}{3f^2}.
\end{equation}

Equation~\ref{eq:EMcl} gives the average EM for a line-of-sight passing through one clump.
Thus to find the EM enhancement due to the clumpy WIM, we multiply this by the expected number of over-densities along a particular line-of-sight
\begin{equation}
N_{{\rm cl}} = \frac{d}{|\sin(b)|L}, \label{e:nclouds}
\end{equation}
where $d$ is the height of the thermal electron content of the Galaxy, and $L$ is the distance between over-densities. For this model, we take $d=1.75$~kpc \citep{Beuermann1985,Gaensler2008}. A better estimate for the electron density scale height may be about 1.4 kpc or even smaller \citep{savagewakker09}. If the scale height would be 1.4 kpc, this would decrease the estimate of EM$_{\rm tot}$ to 20\% since this factor appears only in the estimate of the number of clouds given in eq.~\ref{e:nclouds}.

\subsubsection{Estimating the Mean-Free-Path and Average Dispersion Measure}\label{sec:clumpNumber}
To obtain a value of mean free path, $L$, we use DM$\sin b$, which represents the vertical contribution of electron content only, independent of direction (excepting directions that are very close to the Galactic plane).
The reason for this is that the length of the sight-line and number of clouds are both proportional to $1/\sin(b)$.
The clumpy component of the WIM is DM$_{\rm cl} = \textrm{DM}_{0, {\rm cl}} / \sin b$ and DM$_{\rm cl} \sin(b) = \textrm{DM}_{0, {\rm cl}}$, where DM$_{0, {\rm cl}}$ is the clumpy contribution of dispersion measure through a typical vertical size of the electron layer.
This property can be exploited by using a Poisson distribution for pulsars that possess a DM above a particular value:
\begin{equation}
P(k) = \lambda^k \frac{e^{-\lambda}}{k!},
\end{equation} 
where $P(k)$ is a probability to encounter $k$ clouds along a particular sight-line, and we draw the pulsars' DM value from the ATNF pulsar catalogue\footnote{\url{www.atnf.csiro.au/people/pulsar/psrcat/}} \citep{atnf} with $DM\sin(b)>24.5$~pc~\cmcube, which is taken to infer the pulsar's height above the Galactic plane is higher than the scale height of thermal electrons\footnote{This is in torsion to the value of 16.5~pc~\cmcube given in \citet{Nelemans1997}, and comes from a re-evaluation of pulsar statistics from the ATNF database in the intervening 17 years.}.

The number of clouds along a particular line-of-sight is then $N_{\rm cl}=\lambda = D_p/L$, where $D_p$ is the distance to pulsar and $L$ is the mean path-length between the clumps, was shown above to be the same for all pulsars.
If a sight-line does indeed encounter a clump, then, on average, the DM thus becomes $\mathrm{DM} + \langle \textrm{dm}_1\rangle$.
If we let $F_\mathrm{p}(\lambda)$ be a function which generates a number distributed according to a Poisson distribution with the parameter $\lambda$, one can model pulsars with large DM$\sin b$ as:
\begin{equation}
\textrm{DM}\sin(b) = 24.5\sin(b) +  [\langle \textrm{dm}_1\rangle\sin(b)]F_\mathrm{p}(\lambda).
\end{equation} 

It is possible to compare this value with the observed distribution by means of the Kolmogorov-Smirnov test.
In point of fact, it is not so important that in many cases the additional DM due to the clumpy WIM is not equal to $\langle \textrm{dm}_1\rangle$, because $F_\mathrm{p}(\lambda)$ is small (i.e., of order unity), and hence the difference between distributions according to the Kolmogorov-Smirnov test would not be large.
Additionally, if $F_\mathrm{p}(\lambda)\gtrsim1-2$, then $\langle \textrm{dm}_1\rangle$ is a good estimate of their mean.

Using the Kolmogorov-Smirnov test on 12 pulsars with the highest derived DM$\sin(b)$-value, we find a best estimate of $\langle \textrm{dm}_1\rangle$ and $\lambda$. 
Our method is not particularly sensitive to $\lambda$, and hence our analysis can only restrict the number of clouds to $\lambda\in[2, 4]$.
It is, however, much more sensitive to $\langle \textrm{dm}_1\rangle$, which we restrict to $\langle \textrm{dm}_1\rangle\in[3.6, 3.8]$ pc cm$^{-3}$; almost identical to that from \citet{Nelemans1997}. 
Our derived value of $L/R_c\sim20$ does, however, seem to be much larger than that found in \citet{Nelemans1997}.
We obtain $\lambda = 2-4$, which implies that $L \equiv  d/\lambda= 0.87 -0.43$~kpc.

\begin{figure}
\centering
\includegraphics[width=0.5\textwidth]{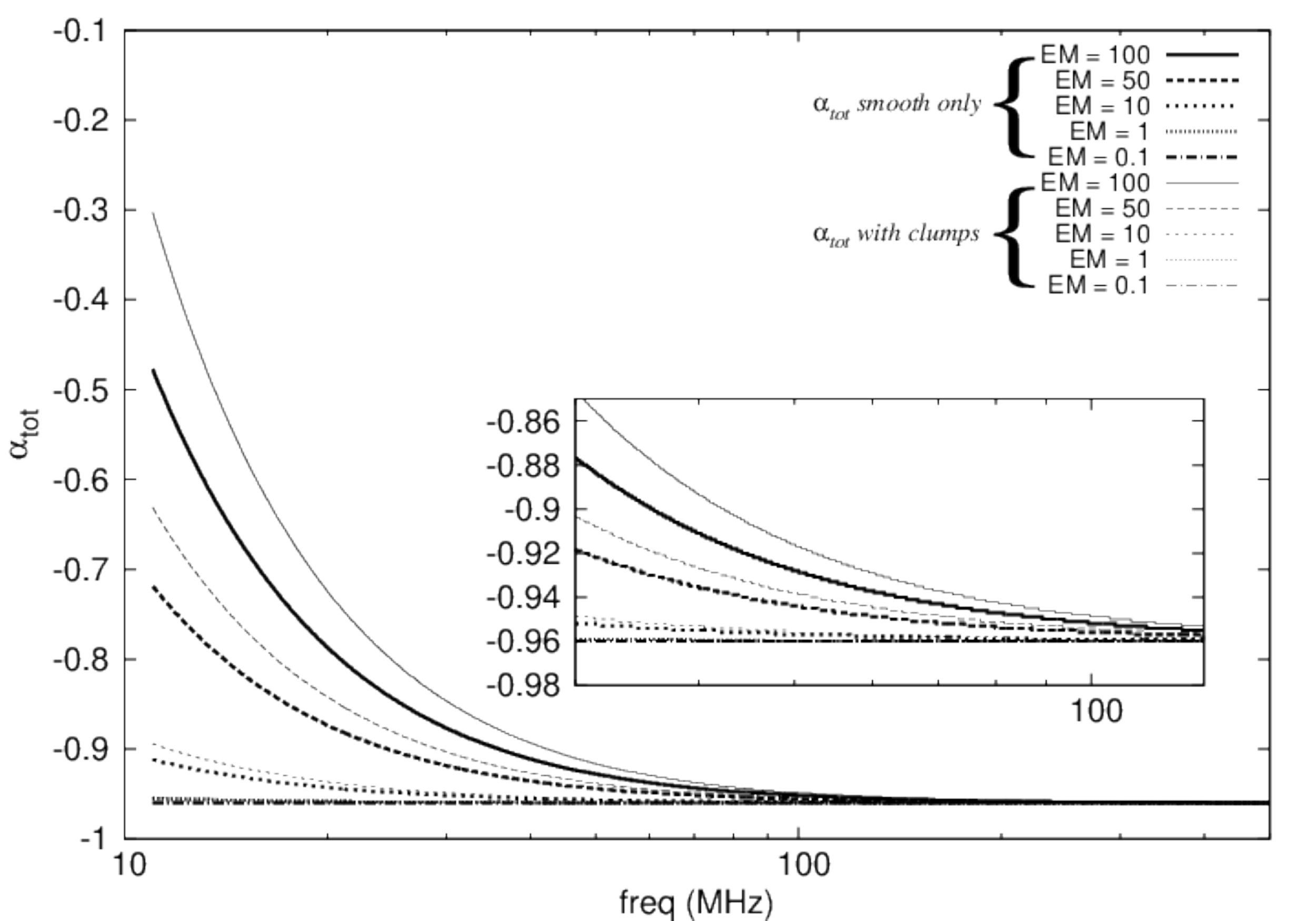}
\caption{A plot of the same {\it total} spectral index (i.e., $\alpha_{\rm tot}=\alpha_{\rm int}+\alphaem)$ as a function of frequency for the same EMs as in Figure~\ref{fig:STAP} (thick lines for EMs as labelled), however, now we have added the clumpy component as modelled in Equation~\ref{eq:EMtot}, assuming a line-of-sight of $b=45^\circ$ (thin lines according to the EM as labelled).
Again, we assume a canonical spectral index of $\alpha_{int}=-0.96$, and calculate a spectral index at 10~MHz intervals (i.e., Figure~\ref{fig:STAP}(b), thick lines).
This plot has the same range in frequency and spectral index as Figure~\ref{fig:STAP}(a) and (b) for easy comparison.
The inset shows a zoom of the frequency region between 30 and 130~MHz.
}
\label{fig:STAP_with_cl}
\end{figure}

\subsubsection{A New Model For the WIM}
Taking into account the clumpy nature of the WIM, one can model STAP as a change in the emission measure due to the smooth and clumpy WIM components:
\begin{equation}\label{eq:EMtot}
\textrm{EM}_{{\rm tot}} = \textrm{EM}_{{\rm sm}} + \textrm{EM}_{{\rm cl}} = \textrm{EM}_{{\rm sm}} + \frac{4N_{{\rm cl}}\langle n_e^2 \rangle R_c}{3f^2},
\end{equation}
where EM$_{{\rm sm}}$ is the smooth component of the WIM and EM$_{{\rm cl}}$ is the emission measure due to the clumpy component, and $D=d/\sin(b)$ is the distance.
Figure~\ref{fig:EM}(b) shows the results of the sum of the two components as in Equation~\ref{eq:EMtot}.
Overall, the EM-sky qualitatively shows the same, stratified structure, but with an increased overall normalisation due to the clumps, with total EMs over significant parts of the sky in the range of 1 to 1000~\emunits.
Figure~\ref{fig:STAP_with_cl} shows the change in total spectral index where, unlike Figure~\ref{fig:STAP}(a) and (b) which only calculate the smooth component of $\alphaem$, the clumpy part of the WIM has been added, assuming a line-of-sight through $b=45^\circ$.
%
%
This figure shows that
the spectral index when such a clumpy distribution is included is changed, comparative to that for a simple smooth distribution.
Quantifying this, at 50~MHz for a source towards which there exists an EM of 100~\emunits and calculating the spectral index in steps of 10~MHz, there is a difference in spectral index of 0.012 between the smooth-only model, and the smooth+clumpy model.
This is a measurable difference between the models, and the smooth+clumpy model gives a spectral index at this frequency of -0.916; a change of 0.05.
Even at frequencies of 110~MHz, which is the lower-end of the LOFAR high band (HBA), there is a difference in spectral index in this scenario of 0.01, suggesting that at least for strong sources, a careful spectral analysis may be able to mesure the thermal electron content of our Galaxy in those source directions.

\subsubsection{Model EM source distribution and application to \halpha data}\label{sec:halpha}
\begin{figure*}
\centering
\includegraphics[width=.9\textwidth]{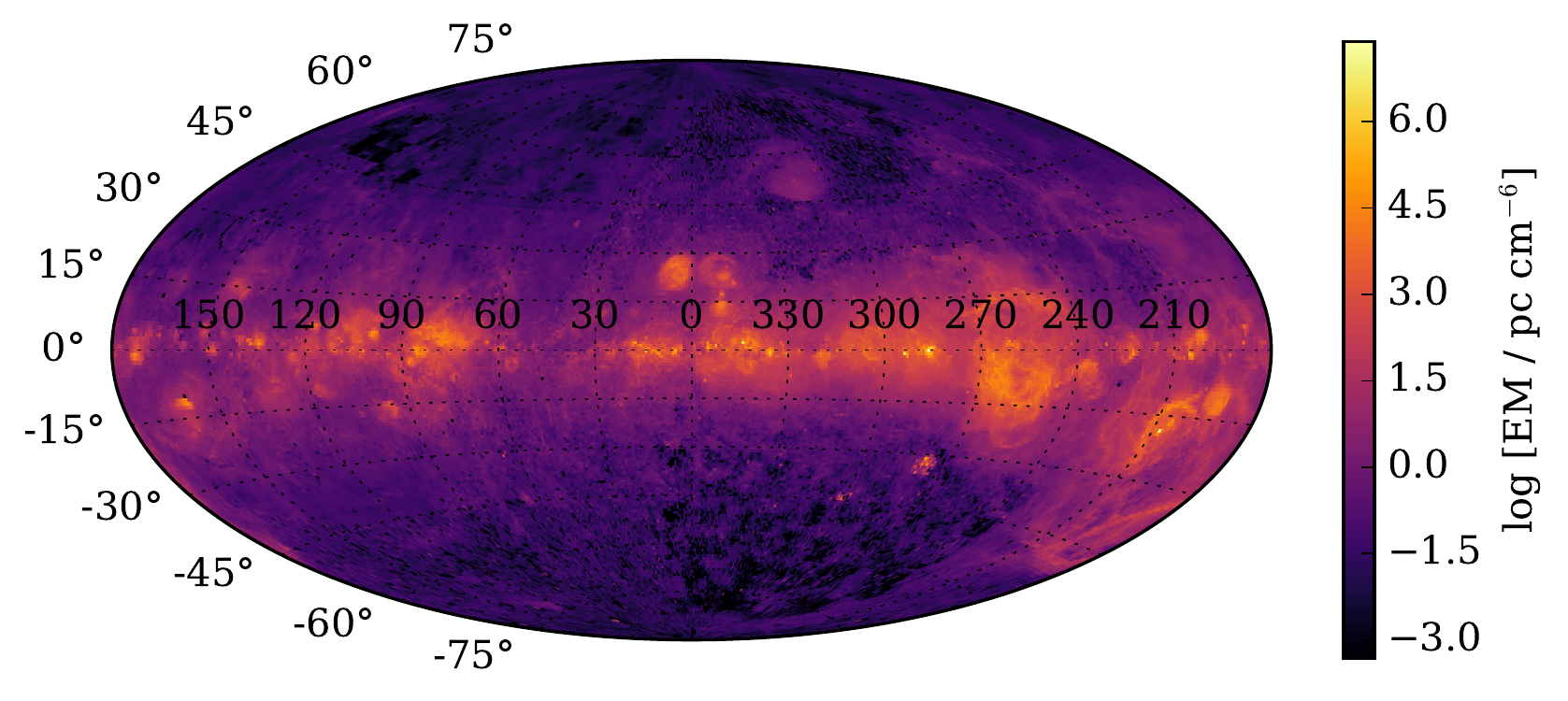}
 \caption{All-sky observed \halpha map based on the WHAM \citep{haffneretal2003} and SHASSA\citep{gaustadetal2001} surveys.}
\label{fig:observedhalpha}
\end{figure*}

\begin{figure}
\centering
\includegraphics[width=0.5\textwidth]{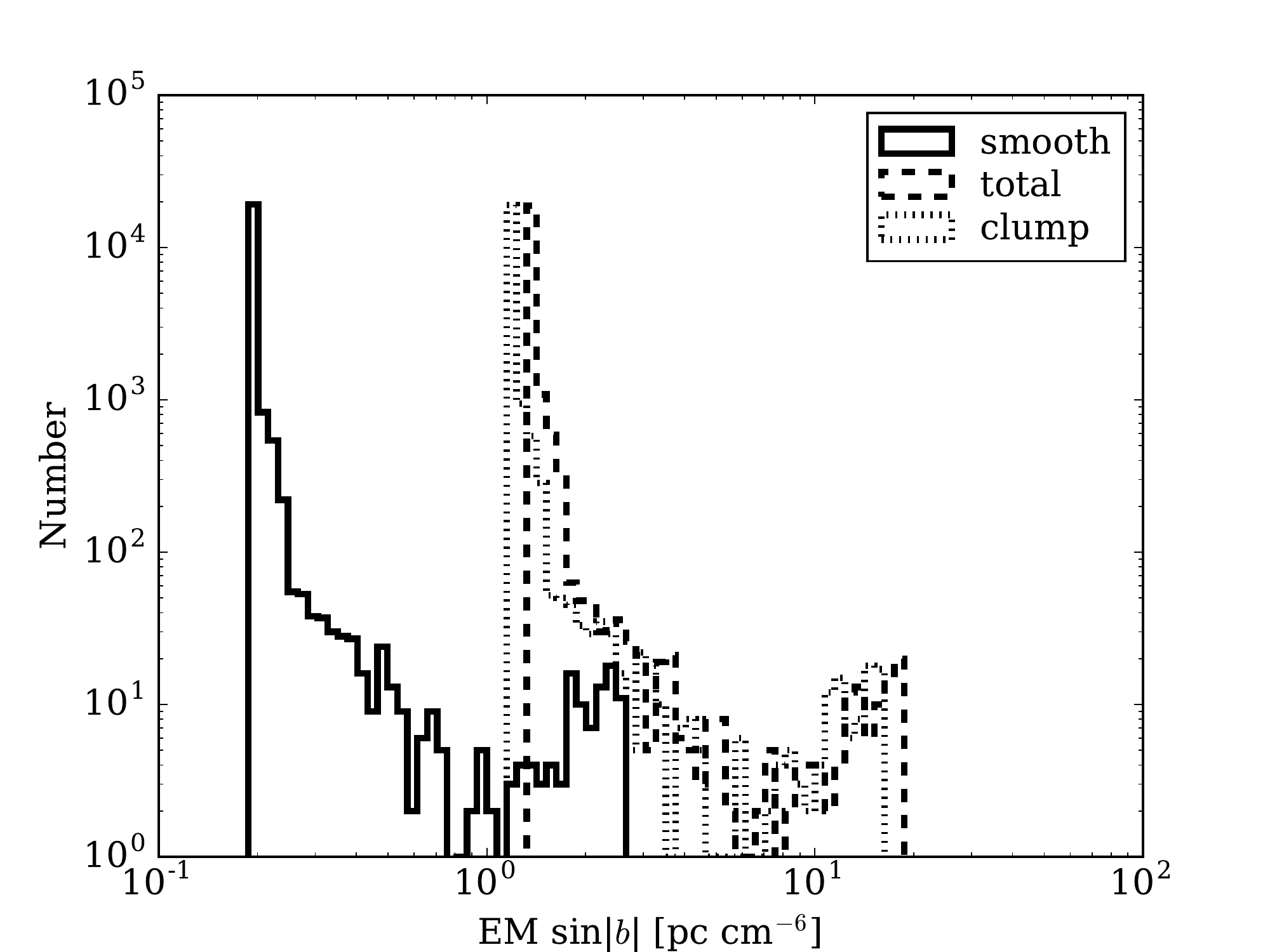}
 \caption{Histogram plot of the EM$|\sin b|$ distribution for sources in the \citet{Farnes2014} meta-catalogue obtained by excluding all sources with $|b|<10^\circ$ and obtained using $f=0.1$, as well as $L=0.43$~kpc.
}
\label{fig:hist}
\end{figure}

It is instructive to compare our model to \halpha data because it is a measure of EM that is independent to our method.
%
However, direct comparison between our EM map and an observed \halpha map would only be possible at high latitudes. At the intermediate and low latitudes, the \halpha emission is so absorbed that the emission represents only the local galaxy (the nearest few kpc). Also, due to the simplicity of our model, it does not contain the individual H~{\sc II} regions that dominate the observed \halpha maps at low latitudes. The \halpha map displayed in Figure~\ref{fig:observedhalpha} demonstrates this point: the map is dominated by local, individual H~{\sc II} regions. However, the general magnitude of the EM distributions between the modeled and observed maps agrees qualitatively. Therefore, we have analysed the EM-distribution (specifically, the EM$|\sin(b)|$-distribution) of our smooth, clumpy and total (smooth+clumpy) models, the results of which are shown in Figure~\ref{fig:hist}.

Relating Figure~\ref{fig:hist} to that of Figure~3 of \citet{Hill2008} demonstrates a number of things.
Firstly, it shows that the distribution of EM for the smooth model does not match that observed in the data, with a peak that is far too small; EM$|\sin(b)|\sim0.1$~\emunits, where the data suggests it should be closer to EM$|\sin(b)|\sim1.4$~\emunits.
This demonstrates that a model for the clumpy nature of the WIM is required for our model to accurately represent the EM-sky.
Secondly, the clumpy model that we derived in Section~\ref{sec:clumpyWIM} is a reasonable fit for the data presented in \citet{Hill2008}.
Figure~\ref{fig:hist} shows that the value of the peak in EM for the total mode (smooth + clumpy) at  EM$|\sin(b)|\sim1.4$ \emunits~almost exactly matches that found by \citet{Hill2008} (see their Table~2). Our distribution hints at a lognormal distribution, however, with a slight overdensity at high EM$|\sin(b)|$.
This suggests that our model is a good {\emph statistical} approximation for the actual EM-sky.
%

\subsection{Consistency check of DMs towards Galactic Globular Clusters}
\begin{figure}
\centering
\includegraphics[width=0.45\textwidth]{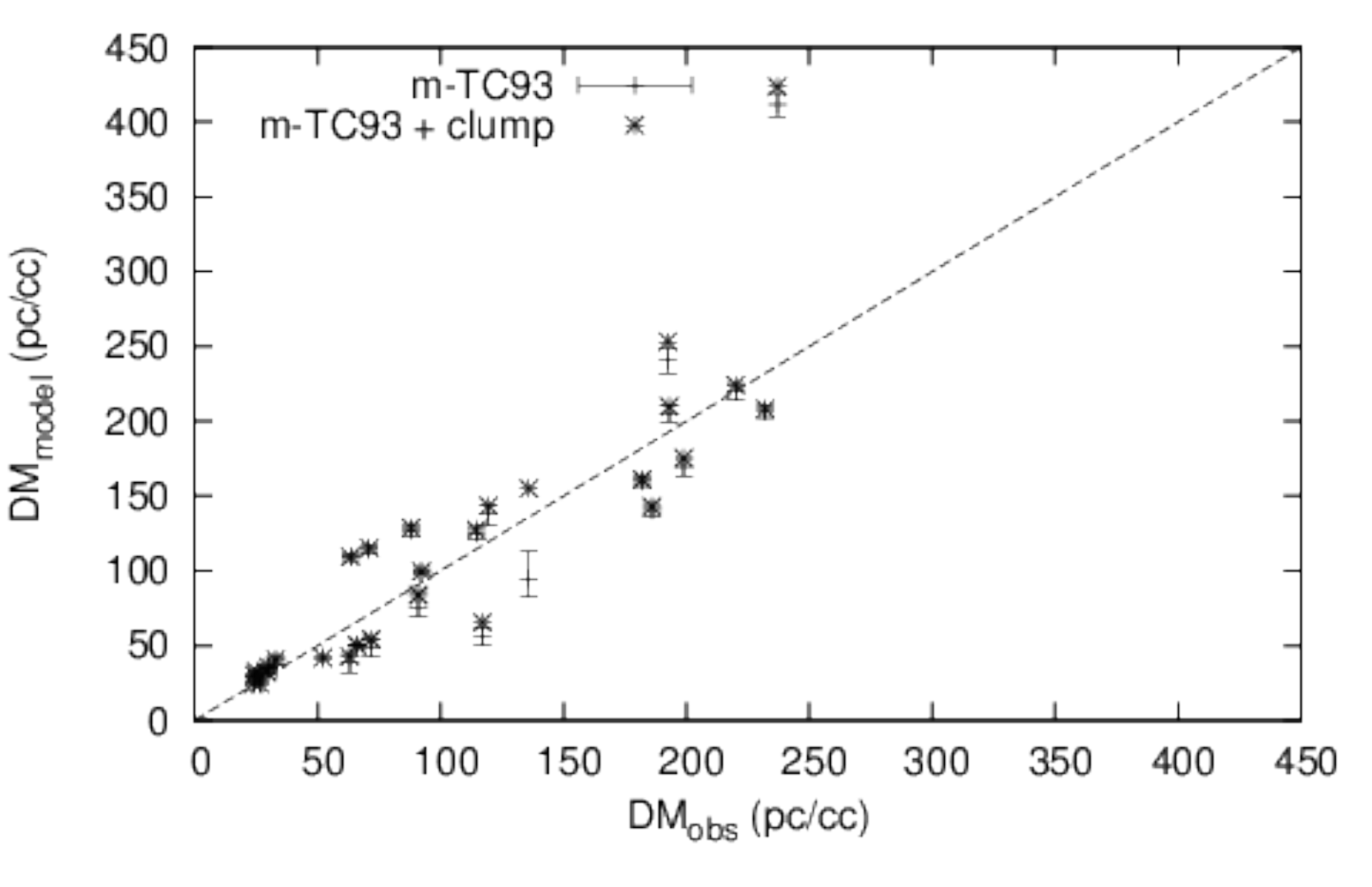}
\caption{A comparison of the observed DMs (in pc~\cmcube: pc/cc) from Globular clusters against that derived from our modified-TC93 model \citep{Schnitzeler2012} (plus-signs) and the modified-TC93 code plus the ``clumpy'' WIM model from Equation~2 of \citet{Nelemans1997} (star-signs).
}
\label{fig:DMs}
\end{figure}

As a consistency check that our method of determining the EM does not also significantly change the DM towards a large sample of pulsars with independently-derived DMs, we have checked the DM that our model predicts against a known pulsar DM population.
Since a precise knowledge of the distance is critical in determining the DM, we require test sources that have well-known distances as well. Therefore, we have chosen pulsars observed in Globular clusters (GCs), since the distances towards the individual GC in question can be determined by independent means.
We have obtained the position, distance to (assuming an error of 0.5~kpc) and average DMs of the pulsars found in 28 GCs from the catalogue of Paulo Freire\footnote{The web-page and relevant references can be found at \url{http://www.naic.edu/~pfreire/GCpsr.html}} to estimate the DM towards these pulsars.
Figure~\ref{fig:DMs} shows the results of this analysis -- the observed DM versus our predicted DM -- for the modified-TC93 only (plus signs) and modified-TC93 and our clumpy model (star-signs), with the dashed line showing a linear relationship between the observed and predicted DMs.
It should be noted that the datum near DM$_{model}\sim420$~pc~\cmcube is for the GC Terzan~5, which is located at $(l,b)=(3^\circ.84,1^\circ.69)$ at a distance of 10.5~kpc, which places it near the Galactic centre, and possibly explaining why it does not follow the DM$_{model}\sim \textrm{DM}_{obs}$ trend that the other data do.
This shows that the DMs towards these regions are not significantly altered by our analysis, and hence that we can be confident that our model does not lead to significantly altered DMs.

\subsection{Source counts and detecting the effect of the WIM}

It was shown above that in our Galaxy the STAP will likely be observable for many parts of the Galaxy, particularly at relatively low latitudes of $|b|\lesssim15^\circ$. Below, we estimate the required accuracy of low-frequency measurements to detect the STAP for a single source. Also, we calculate the required source density at higher latitudes which would enable a statistical detection of STAP in an ensemble of sources.

\subsubsection{Detection of the STAP in a single source}

\begin{figure}
\centering
\includegraphics[width=0.5\textwidth]{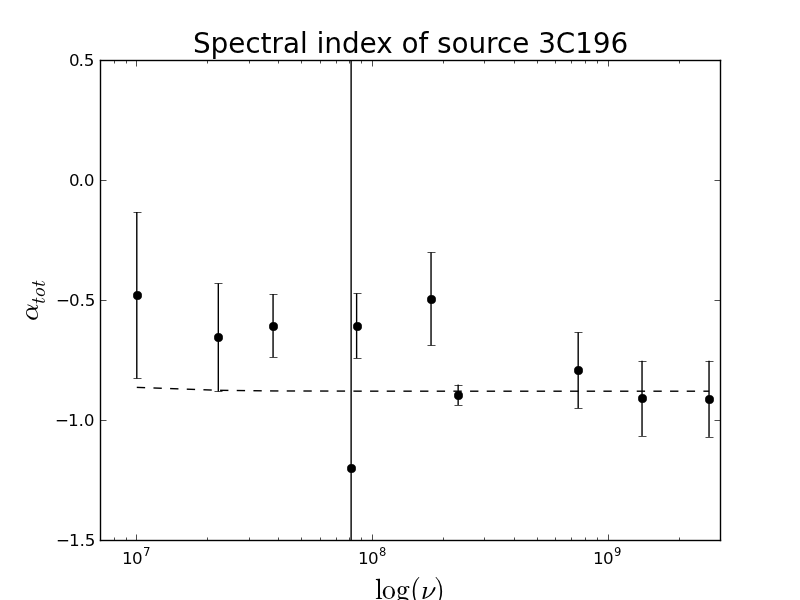}
\includegraphics[width=0.5\textwidth]{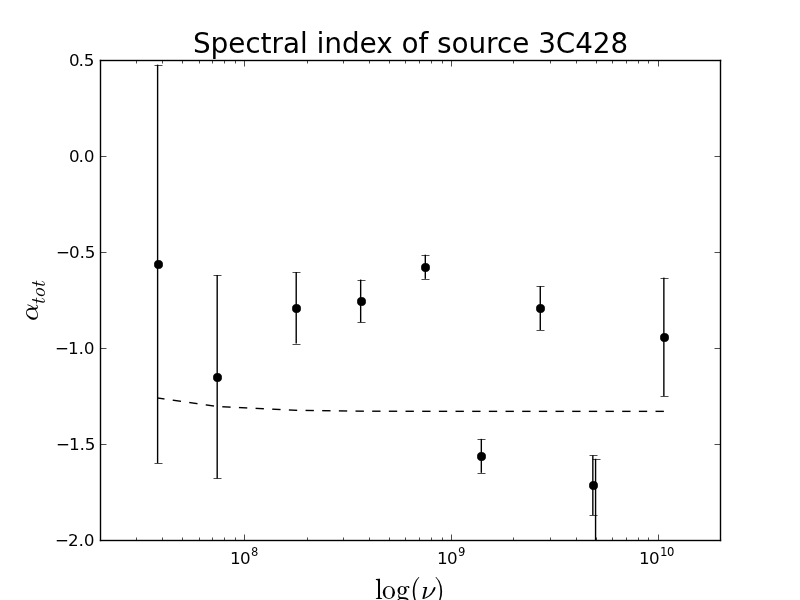}
\caption{Total spectral index $\alpha_{\rm tot}$ as a function of frequency for 3C196 (top) and 3C428 (bottom). The overlaid dashed line is the calculated  $\alpha_{\rm tot}$ based on expected EM and intrinsic spectral index, as discussed in the text.}
\label{fig:examples}
\end{figure}

Figure~\ref{fig:STAP_with_cl} shows that very low-frequency observations are required for a detectable deviation of the measured spectral index from the intrinsic one. In figure~\ref{fig:examples}, we the total spectral index $\alpha_{\rm tot}$ as calculated according to eq.~(\ref{eq:alpha2}) for two extragalactic sources. The left plot shows data from 3C196 \citep{Kellermann1969,scaifeheald2012}, a strong source often used as calibrator source, which was chosen for the wealth of available observations. The right plot depicts observations from 3C428 (obtained from the NED database), which is located at low Galactic latitude ($b = 1.31^{\circ}$) so that the effect of the Galactic WIM would be higher. The dashed lines show the intrinsic spectral index ($-0.88$ for 3C196 and $-1.33$ for 3C428 \citep{Kellermann1969}) at the higher frequencies and the expected change in spectral index based on the EM obtained from our modified-TC93+clumpy model.

However, the current low-frequency instruments LOFAR and MWA are planning or executing surveys at sufficiently low freqencies to etect this effect in principle in a single source. As an example, we simulated detectability of a Galactic EM = 100~pc~cm$^{-6}$ in the ``Deep'' survey (Tier 2) of the currently ongoing ``Tier surveys'' with LOFAR \citep{rottgeringetal2011}. For the planned 30~MHz survey, assuming 1~MHz frequency channels over the bandwidth of 16~MHz, we fit equation~(\ref{eq:alpha2}) as a function of frequency channel. A simple monte carlo analysis shows that this EM would be detectable at the $5\sigma$ level in a source with flux $S_{30~\mbox{MHz}} = 7$~Jansky.

\subsubsection{Statistical detection of the STAP in an ensemble of sources}

We explore here how the sensitivity of next-generation radio telescopes such as LOFAR, MWA and especially the SKA may allow the  statistical detection of this phenomenon at higher latitudes, where the low EM towards sources make STAP analysis currently difficult or impossible.
At latitudes higher than $b\sim30^\circ$, Figure~\ref{fig:EM}(b) shows that an average of EM$\sim0.1-1$~\emunits, which Figure~\ref{fig:STAP_with_cl} shows a change in spectral index at 30~MHz of $\sim0.01$.
Given that \citet{Randall2012} showed that the spectral index distribution of sources between 843~MHz and 1.4~GHz is Gaussian with a standard deviation of $\sigma_\alpha=0.56$ (after K-correction) for an average spectral index of $\alpha_{\rm meas}=-0.5$, one can detect a change in the distribution that is statistically significant of:
\begin{equation}
\Delta\sigma\gtrsim\frac{\sigma_{\alpha}}{\sqrt{N}},
\end{equation}
where $N$ is the number of sources per square degree.
For a source density of 10, 100 and 1000 sources per square degree, then, one can robustly detect a change in spectral index of $\alphaem\sim0.18,0.06$, and 0.02.
This shows that, for source densities of 1000 sources per square degree, the thermal electron distribution at latitudes higher than $b\sim30^\circ$ could be (statistically) explored.
Thus, sensitive, dedicated low-frequency surveys with telescopes such as LOFAR and MWA and eventually the SKA should be able to probe this effect and significantly enhance our knowledge of the thermal electron distribution of the Galaxy.

\section{Conclusions}
In this paper, we have presented a new method for exploring the thermal electron density of our Galaxy by means of spectral analysis of background sources in the limit of a zero-source-function solution to the equation of radiative transfer.
We have shown the following:

\begin{enumerate}
\item This limit results in an additive term to the intrinsic spectral index, if said index is calculated using natural logarithms.
This effect is particularly important at low frequencies, and especially-so at frequencies below $\sim100$~MHz, even for modest emission measures of between 1 and 100~\emunits.

\item Through the use of the meta-catalogue of \citet{Farnes2014} (used as a position-template for possible low-frequency point source catalogs), we find that these EM values are representative of large parts of the Galaxy.

\item A smooth distribution of thermal electrons does not represent the known EM distribution obtained from \halpha data from the WHAM survey, which is an independent measure of the EM.

\item A simple model for the clumpy WIM, on the other hand, does indeed produce such a distribution, and increases the normalisation of EMs over the sky to 100-1000~\emunits, at least at relatively low Galactic latitudes (i.e., $|b|\lesssim15^\circ$).
However, we also showed that because of this higher normalisation that the clumpy WIM gives, it should be possible to explore the Galactic distribution of thermal electrons at higher Galactic latitudes statistically with source counts of $\sim1000$ sources per square degree.
\end{enumerate}

The latter two items confirm earlier work based on H$\alpha$ observations \citep{Hill2008} and pulsar dispersion measures \citep{berkhuijsenfletcher2008}, using lognormal distributions.

Given that an excellent knowledge of the thermal electron distribution is crucial for our knowledge of the magnetic field strength and structure in our Galaxy, which is itself an integral part of the Galactic ecology, this new method, which is independent of rotation measures but can be observed {\it at the same frequencies}, could make an important contribution to our knowledge of our magnetic Galaxy.

\section*{Ackowlegements}DIJ would like to acknowledge enlightening conversations with Dominic Schnitzeler (who also kindly provided his modified-TC93 code), Bryan Gaensler, Jamie Farnes, Marco Iacobelli, Cameron van Eck, and Elmar Koerding. This research has made use of the NASA/IPAC Extragalactic Database (NED) which is operated by the Jet Propulsion Laboratory, California Institute of Technology, under contract with the National Aeronautics and Space Administration. MH acknowledges the support of research programme 639.042.915, which is partly financed by the Netherlands Organisation for Scientific Research (NWO). The authors acknowledge use of the Southern H-Alpha Sky Survey Atlas (SHASSA), which is supported by the National Science Foundation.

\label{lastpage}
\bibliographystyle{mnras} 
\bibliography{bibliography} 

\end{document}